\documentclass[9pt,amsmath,amssymb,nofootinbib]{revtex4}
\usepackage{amsmath,amsfonts}
\usepackage{graphicx}
\usepackage{hyperref}
\newcommand{\be}{\begin{equation}}
\newcommand{\ee}{\end{equation}}

\newcommand{\matrice}{\begin{pmatrix}}
\newcommand{\ematrice}{\end{pmatrix}}
\newcommand{\bea}{\begin{eqnarray}}
\newcommand{\eea}{\end{eqnarray}}

 \let\b=\beta  \let\g=\gamma  \let\d=\delta
     \let\th=\theta \let\k=\kappa 
\let\m=\mu    \let\n=\nu              \let\om=\omega
\let\s=\sigma      
    
 \let\D=\Delta    
         
  \let\eps=\epsilon

\let\==\equiv

\let\p=\partial

\def\FF{{\cal F}} 
 
\def\RR{{\cal R}}  
 \def\SS{{\cal S}}

\def\to{\rightarrow}

\def\ra{\right\rangle}

\def\ra{\rangle}

\begin{document}
\title{Rainbow statistics}
\author{Michele Arzano}
\email{marzano@perimeterinstitute.ca}
\affiliation{Perimeter Institute for Theoretical Physics\\
31 Caroline St. N, N2L 2Y5, Waterloo ON, Canada}
\author{Dario Benedetti}
\email{dbenedetti@perimeterinstitute.ca}
\affiliation{Perimeter Institute for Theoretical Physics\\
31 Caroline St. N, N2L 2Y5, Waterloo ON, Canada}

\begin{abstract}
\vskip .2cm
\begin{center}
{\bf Abstract}
\end{center}
Non-commutative quantum field theories and their global quantum group symmetries provide an intriguing attempt
to go beyond the realm of standard local quantum field theory.  A common feature of these models is that the
quantum group symmetry of their Hilbert spaces induces additional structure in the multiparticle states which
reflects a non-trivial momentum-dependent statistics.  We investigate the properties of this ``rainbow
statistics" in the particular context of $\kappa$-quantum fields and discuss the analogies/differences with
models with twisted statistics.

\end{abstract}

\maketitle


\section{Introduction}
A fundamental postulate of quantum mechanics asserts that the physical states of a quantum system are represented by rays in a complex Hilbert space. In relativistic quantum theory the ``one-particle" Hilbert space for a scalar boson can be obtained from a classical scalar field equipping the space of solutions of the Klein-Gordon equation with a {\it compatible complex structure} and restricting to an appropriate (positive energy) subspace \cite{Ashtekar:1975zn}.  In non-commutative field theory the space of solutions of the classical equations of motion is turned into a non-commutative algebra and such ``quantization" procedure is often justified by the desire to include effects due to quantum space-time uncertainty \cite{Doplicher:1994zv}.  One approach to non-commutative field theory (NCFT) relates the emergence of a non-commutative structure to a mathematical well defined notion of quantization of the algebraic structure of the isometries of the ambient space on which the fields live \cite{Chaichian:2004za, Majid-Ruegg}.  Such point of view besides revealing a richer and consistent mathematical framework underlying NCFTs also suggests that non-commutative quantum field theory (NCQFT) might provide a natural extension of ordinary local quantum field theory when the requirement of locality for observables is mildly relaxed \cite{Arzano:2007nx}. Under this new light NCQFTs become much more than just ``deformations" of usual quantum field theories whose new features are expected to emerge only in planckian regimes.  They provide an example of consistent models incorporating non-local effects which in the undeformed limit reduce to ordinary local QFT.  The new features they exhibit might play a key role in the understanding of phenomena in which local QFT as an effective theory is expected to fail like e.g. 
in extreme conditions on non-trivial background geometries when the requirement of locality appears to be too stringent (see e.g. \cite{Lowe:1995ac, Giddings:2006sj}).\\
The novel algebraic structures which appear in the description of the relativistic symmetries of non-commutative fields are non-trivial Hopf algebras also known as {\it quantum groups}.  Twisted and $\kappa$-deformed Poincar\'e algebras are the most popular examples of such Hopf algebras and are related, respectively, to fields on the Moyal plane \cite{Chaichian:2004za, akofor} and on $\kappa$-Minkowski non-commutative space-time \cite{Majid-Ruegg}.  As it will be discussed in detail below the non-trivial co-algebra sector of these algebras affects the multiparticle sector of their respective quantum field theories.  From a representation theoretic point of view this is due to the non-trivial nature of the intertwiners between tensor products of one-particle Hilbert spaces which are themselves irreducible representations of the deformed algebras.  This more involved state of affairs is reflected in the {\it momentum-dependent statistics} of multi-particle states.  In the present paper we attempt to provide a unified view of such ``rainbow statistics" with special focus on the case of the $\kappa$-Poincar\'e algebra\footnote{This particular deformation of the Poincar\'e algebra has received much attention in recent years for its possible role as the algebra of generators of relativistic symmetries preserving an observer independent energy scale, $\kappa$, often identified with the Planck energy \cite{AmelinoCamelia:2000mn}.}.  Early work on the construction of quantum field theories with such deformed symmetries focused on perturbative path integral approaches \cite{AmelinoCamelia:2001fd}.  More recently, motivated by the search for a quantum version of the Noether charges associated with $\kappa$-symmetries derived in a classical field theory setting \cite{Agostini:2006nc}, the focus shifted on canonical quantization.  The construction of a $\kappa$-deformed Fock space \cite{AM-fock} led to the emergence of a momentum-dependent statistics for which, despite the efforts of different research groups, there is so far only a partial understanding.\\
In the next section we review the basics of statistics in ordinary quantum field theory emphasizing the role that (trivial) Hopf algebraic structures have even in this simple case.  In section III we describe the emergence of deformed statistics in the well studied case of quantum fields on the Moyal plane as a warm up for the following discussion on $\kappa$-statistics.  Section IV contains the main results of this work namely an extension of the deformed bosonization given in \cite{AM-fock} to include massive scalar fields formulated in a way which generalizes to any basis of the $\kappa$-Poincar\'e algebra and our insights on the problem of covariance of $\kappa$-Fock spaces.  We conclude in section V with some remarks on what appears to be the status of $\kappa$-deformed statistics in light of the results of our discussion.


\section{Statistics in local quantum field theory}
Bosonic and fermionic behaviors are first observed in non-relativistic quantum mechanics where the statistical properties of quantum point particles are deeply connected with the topology of their classical configuration space.  Indeed inequivalent quantizations of a classical system are labeled by the unitary irreducible representations of the fundamental group of the configuration space \cite{BalMarmo}. The symmetric group $S_n$ is always a subgroup of the fundamental group for a system of $n$ identical particles in $d\geq 3$ dimensions, hence Bose-, Fermi- and para- statistics arise as inequivalent quantizations. Different statistics are also possible if particles have a nontrivial internal structure, as for example in the case of the cyclic statistics of \cite{surya}.  For $d=2$ the braid group $B_n$ takes the place of $S_n$ giving rise to braid (or anyonic) statistics \cite{Wu}.\\
In local quantum field theory the Fock space is built from direct sums of symmetrized (anti-symmetrized) tensor products of one-particle Hilbert space for bosons (fermions).  These constructions are based on the action of the trivial and antisymmetric representations of the permutation group.  Indeed these ``exchange operators" on tensor products of Hilbert spaces are intertwiners for the Poincar\'e algebra and the result of their action on the Hilbert space has no physical consequence.\\  
We focus from now on on the bosonic Fock space of a scalar field.  As mentioned above such space carries a trivial representation of the symmetric group.  This is the discrete group of permutations of $n$ objects and is denoted by $S_n$.  The symmetric group $S_n$ has a presentation in terms of a set of $n-1$ generators $s_i$ for $i=1,...,n-1$ satisfying the relations
\begin{eqnarray} \label{symgroup1}
&s_i s_j = s_j s_i &\text{for}\  |i-j|>1\ ,\\ \label{symgroup2}
& s_i s_{i+1} s_i = s_{i+1} s_i s_{i+1} &\text{for}\  1\leq i\leq n-2\ ,\\ \label{symgroup3}
& s_i^2 =1 &\text{for}\ 1\leq i\leq n-1\ .
\end{eqnarray}
In the natural permutation representation the generator $s_i$ acts on an $n$-dimensional vector space $V$ by exchanging the $i$-th and the $i+1$-th component
of a vector $\vec x \in V$, $i.e.$ $\pi_0(s_i)\cdot (x_1, ... x_i, x_{i+1},...x_n)=(x_1, ... x_{i+1}, x_i,...x_n)$.
This representation, like any permutation representation, is a reducible one.
Standard representation theory (see for example \cite{fulton}) teaches us how to classify the irreducible representations, and as it turns out these are classified by Young tableaux.\\
The action of an element of the symmetric group in quantum field theory is represented by a permutation of the elements of a tensor product of single-particle states
\be \label{perm}
\pi_0(s_i) |\vec p_1\ra \otimes ... |\vec p_i\ra\otimes |\vec p_{i+1}\ra \otimes ...|\vec p_n\ra=
|\vec p_1\ra \otimes ...
|\vec p_{i+1}\ra\otimes |\vec p_i\ra \otimes
 ...|\vec p_n\ra\ .
\ee
As a consequence of the spin-statistics theorem physical $n$-particle states $|\vec p_1,...p_n\ra$ for identical particles have to be in the trivial representation
of the symmetric group for the case of bosons, $i.e.$
\be
\pi_0(s_i) |\vec p_1,...\vec p_n\ra=|\vec p_1,...\vec p_n\ra\ ,
\ee
and in the alternating one for the case of fermions, $i.e.$
\be
\pi_0(s_i) |\vec p_1,...\vec p_n\ra=-|\vec p_1,...\vec p_n\ra\ .
\ee
The case of non abelian (multi-dimensional) unitary irreducible representations has also been considered in the literature, and is known as parastatistics \cite{Green}.  Keeping in mind that a one-particle Hilbert space, as a complex vector space, is a an irreducible representation of the Poincar\'e group, the collection of such spaces together with the tensor product and the (trivial) action of the symmetric group constitutes what in mathematical lingo is known as {\it symmetric monoidal category} (see \cite{Oeckl:2005rh} for an accessible introduction to the basic concepts of category theory).  This category point of view will be useful later as it shows how more general statistics structures can arise when one moves from the category of representations of Lie groups to quantum groups.\\
A general way to reduce a representation like (\ref{perm}) to the one-dimensional irreducible one of bosons is by acting on the un-symmetrized $n$-particle states $|\vec p_1\ra\otimes|\vec p_2\ra...\otimes\vec |p_n\ra$ with the operator 
\be \label{symm}
\SS^{(n)}=\frac{1}{\sqrt{n!}}\sum_{\{j_w\}\in W(\pi_0(s_1),...\pi_0(s_{n-1}))}\prod_{j_w}\pi_0(s_{j_w})\ ,
\ee
where the sum is over the set $W(\pi_0(s_1),...\pi_0(s_{n-1}))$ of all possible words (modulo the relations (\ref{symgroup1}) to (\ref{symgroup3})) made with the operators $\pi_0(s_i)$ (it is well known that there are in total $n!$ of such words).
Given the one-particle Hilbert space for our scalar quantum field  $\mathcal{H}$, we can construct the symmetric Fock space $\mathcal{F}_s(\mathcal{H})$ as follows, denoting
$\mathcal{H}_n=\mathcal{H}^{\otimes n}$ (with $\mathcal{H}^0=\mathbb{C}$) the (symmetric) bosonic Fock space is given by
\begin{equation}
\mathcal{F}_s(\mathcal{H})=\bigoplus_{n=0}^{\infty}\SS^{(n)}\mathcal{H}_n\, .
\end{equation}
In the remaining of the paper we will be interested in quantum deformations of the Poincar\'e algebra and their effects on the structure of the bosonic Fock space.  As we will make extensive use of Hopf algebras we will briefly show here how such objects emerge even at the level of familiar local quantum field theory.\\  Hopf algebras naturally arise in representation theory whenever one tries to build new representations of a group (algebra) from the tensor product of their modules.  In particular this applies when one constructs multi-particle Hilbert spaces from the one-particle Hilbert space carrying an irreducible representation of the Poincar\'e algebra.  For simplicity (and because the basis of modes usually chosen to describe $\mathcal{H}$ are eigenfunctions of translations) we will restrict to the subalgebra of translations spanned by the generators $P_{\mu}$.  The one-particle Hilbert space $\mathcal{H}$ carries a 
representation of the translation algebra $T_4$ and of its universal enveloping algebra $U(T_4)$.  Now consider the ``un-symmetrized" two-particle space $\mathcal{H}_2=\mathcal{H}\otimes\mathcal{H}$, the natural question to ask is how do we define a representation of $T_4$ on it or, in other words, what is the action of $P_{\mu}$ on $|\vec{p}_1\ra\otimes|\vec{p}_2\ra\in \mathcal{H}_2$ ?  We know that the $P_{\mu}$'s act like derivatives so we will {\it define} 
\begin{equation}
\label{leibn}
P_{\mu}\vartriangleright|\vec{p}_1,\vec{p}_2\ra\equiv (P_{\mu}\vartriangleright|\vec{p}_1\ra)\otimes|\vec{p}_2\ra+|\vec{p}_1\ra\otimes (P_{\mu}\vartriangleright|\vec{p}_2\ra)
\end{equation}
In the language of Hopf algebras $U(T_4)$ possesses a new structure, the coproduct, defined by
\begin{equation}
\Delta P_{\mu}=P_{\mu}\otimes 1+ 1\otimes P_{\mu}
\end{equation}
which encodes exactly the action (\ref{leibn})
\begin{equation}
P_{\mu}\vartriangleright|\vec{p}_1,\vec{p}_2\ra\equiv \Delta P_{\mu}\vartriangleright(|\vec{p}_1\ra\otimes|\vec{p}_2\ra)
\end{equation}
without the need of additional assumptions put ``by hand"; the representation theory for tensor product spaces is ``built in" the Hopf algebra structure.\\  Another key fact regarding Hopf algebra structures emerges from the interplay between the representation of the symmetric group and that of translation symmetry on multiparticle states.  Focusing again, for simplicity, on the two-particle state $\mathcal{H}_2$, we note that the element of the permutation group $S_2$ acts as an intertwiner of representations.  This equips $U(T_4)$ with a trivial {\it quasitriangular structure}.  To make the statement clear we briefly recall the notion of quasi-triangular Hopf algebra \cite{Chari:1994pz, Majid}. 
A Hopf algebra $H$ is said to be {\it quasitriangular} if there exists an element $\mathcal{R}=r_{\alpha}\otimes r^{\alpha}\in H \otimes H$ such that for every $a\in H$
\begin{eqnarray}
&\s \Delta(a) =\mathcal{R}\Delta(a)\mathcal{R}^{-1} \label{r-struct1}\\
&(id\otimes\Delta)\mathcal{R}= \mathcal{R}_{13}\mathcal{R}_{12}\,,\,\,\, 
(\Delta\otimes id)\mathcal{R}=\mathcal{R}_{13}\mathcal{R}_{23}  \label{r-struct2}
\end{eqnarray}
where $\sigma$ is the flip map and $\RR_{12}=r_{\alpha}\otimes r^{\alpha}\otimes 1$,  
$\RR_{13}=r_{\alpha}\otimes 1 \otimes r^{\alpha}$, etc.  Now $\pi_0(s_1)$ acting on the state  
$|\vec{p}_1\ra\otimes|\vec{p}_2\ra$ can be written as a composition of the flip operator $\sigma$ and an element $1\otimes 1\in U(T_4)\otimes U(T_4)$ as $\pi_0(s_1)=\sigma \circ (1\otimes 1)$.  One can easily see that the element $\RR=1\otimes 1$ trivially satisfies the properties above.\\ The possibility for $U(T_4)$ to acquire a non-trivial quasi-triangular structure under quantum deformations will be of key relevance in the next sections.
Indeed we will see that in non-commutative quantum field theory the algebra of functions on Minkowski space is turned into a non-commutative algebra through the introduction of a $*$-multiplication.  The new framework radically changes the structures described above, in particular we will see how, according to the type of ``non-commutativity" one can have a non-trivial quasitriangular structure and a deformed action of the translation generators on multiparticle states.  These novel features can have rather interesting consequences for the statistics of the free modes of a scalar field as we will see in the case of $\kappa$-quantum fields.


\section{Twisted statistics on the Moyal plane}
The first level of complexity is encountered with the example of canonical non-commutative spacetime or Moyal plane
\begin{equation}
\label{moyalcomm}
[x_{\mu},x_{\nu}]=i\theta_{\mu\nu}
\end{equation}
with $\theta_{\mu\nu}$ a real antisymmetric constant matrix.  Functions on the Moyal plane form a non-commutative algebra $\mathcal{C}_{\theta}(\mathbb{R}^{3+1})$.  Such algebra can be obtained 
\cite{oeckl-twist} as a deformation of the commutative algebra of smooth functions on Minkowski space $\mathcal{C}(\mathbb{R}^{3+1})$.  As a vector space $\mathcal{C}(\mathbb{R}^{3+1})$ carries a representation of the translation group $\mathbb{R}^{3+1}$.  At the same time $\mathcal{C}(\mathbb{R}^{3+1})$, as the algebra of functions on a group,  has an associated Hopf algebra structure, in particular the coproduct $\Delta f\in \mathcal{C}(\mathbb{R}^{3+1})\otimes\mathcal{C}(\mathbb{R}^{3+1})\simeq
\mathcal{C}(\mathbb{R}^{3+1}\times\mathbb{R}^{3+1})$ encodes the translation group law
\begin{equation}
(\Delta f) (x,y)= f(x+y)\, .
\end{equation}
The pointwise product in $\mathcal{C}(\mathbb{R}^{3+1})$ can be deformed to a non-commutative 
$\star$-product via a ``twist" map
\begin{equation}
\mathcal{F}_{\theta}=\exp{\left(-\frac{i}{2}\theta_{\mu\nu}
\frac{\partial}{\partial x_{\mu}}\frac{\partial}{\partial y_{\nu}}\right)}
\end{equation}
in the following way: if $m: \mathcal{C}(\mathbb{R}^{3+1})\otimes\mathcal{C}(\mathbb{R}^{3+1})\rightarrow \mathcal{C}(\mathbb{R}^{3+1})$ gives the pointwise product
\begin{equation}
m(f\otimes g)(x)=f(x)\cdot g(x)
\end{equation}
the $\star$-product will be given by $m_{\theta}=m\circ\mathcal{F}^{-1}_{\theta}$
\begin{equation}
m_{\theta}(f\otimes g)(x)= \exp{\left(\frac{i}{2}\theta_{\mu\nu}
\frac{\partial}{\partial x_{\mu}}\frac{\partial}{\partial y_{\nu}}\right)}  f(x)\cdot g(y) |_{y=x}\, ,
\end{equation}
the vector space $\mathcal{C}(\mathbb{R}^{3+1})$ endowed with the non-commutative product $m_{\theta}$ becomes the non-commutative algebra $\mathcal{C}_{\theta}(\mathbb{R}^{3+1})$.
The twist can be seen as an element of the tensor product of the universal enveloping algebra of the translation generators $\mathcal{F}_{\theta}\in U(T_4)\otimes U(T_4)$
\begin{equation}
\mathcal{F}_{\theta}=\exp{\left(\frac{i}{2}\theta^{\mu\nu}
P_{\mu}\otimes P_{\nu}\right)}\, ,
\end{equation}
which generate a twisted Hopf algebra $U_{\theta}(T_4)$ with a {\it non-trivial quasitriangular structure}, $i.e.$ with a non-trivial R-matrix
$\mathcal{R} = \mathcal{F}_{\theta}^{-2}$.  To see this in a more concrete way we turn back to the undeformed Hilbert space of a scalar quantum field $\mathcal{H}$.  
After twisting $\mathcal{H}$, seen as an algebra of functions, turns into a non-commutative algebra $\mathcal{H}_{\theta}$ equipped with the twisted product $m_{\theta}=m\circ\mathcal{F}_{\theta}$.  
The coproduct of translation generators  $P_{\mu}\in U_{\theta}(T_4)$ does not change under twisting
\begin{equation}
\label{thetacop}
 \D_{\th}(P_\m)\equiv\mathcal{F}_{\theta}\D(P_\m)\mathcal{F}^{-1}_{\theta}=\D(P_\m)\,.
 \end{equation}
The fact that $P_{\mu}$s act as derivatives on functions provides a natural pairing between $U_{\theta}(T_4)$ and $\mathcal{C}_{\theta}(\mathbb{R}^{3+1})$ (see e.g. \cite{oeckl-twist}).  The product on $\mathcal{H}_{\theta}$ and the coproduct of $U_{\theta}(T_4)$ are related by such pairing through
\begin{equation}
\label{pairprop}
\langle \Delta_{\theta}(P_{\mu})\, , \,(\phi_1\otimes\phi_2)\rangle\equiv\langle P_{\mu}\, ,\, m_{\theta}(\phi_1\otimes\phi_2) \rangle
\end{equation}
We look for a representation of $S_2$ on the ``un-symmetrized" two-particle space $\mathcal{H}_{\theta}\otimes\mathcal{H}_{\theta}$.   A first guess would be to consider the usual $\pi(s_1)=\sigma$ with $\sigma$ is the flip operator.  Notice however that if in (\ref{pairprop}) we use (\ref{thetacop}) and act with $\pi(s_1)$ on the tensor product we have
\begin{equation}
\langle \Delta(P_{\mu})\, , \,\pi(s_1)(\phi_1\otimes\phi_2)\rangle\equiv\langle P_{\mu}\, ,\, m_{\theta}(\pi(s_1)(\phi_1\otimes\phi_2)) \rangle
\end{equation}
we have for the left hand side
\begin{equation}
\langle \Delta(P_{\mu})\, , \,\sigma(\phi_1\otimes\phi_2)\rangle=
\langle \sigma\circ\Delta(P_{\mu})\, , \,(\phi_1\otimes\phi_2)\rangle=
\langle\Delta(P_{\mu})\, , \,(\phi_1\otimes\phi_2)\rangle
\end{equation}
while for the right hand side
\begin{equation}
\langle P_{\mu}\, ,\, m_{\theta}(\sigma(\phi_1\otimes\phi_2)) \rangle=
\langle P_{\mu}\, ,\, m_{\theta}(\phi_2\otimes\phi_1) \rangle\neq 
\langle P_{\mu}\, ,\, m_{\theta}(\phi_1\otimes\phi_2) \rangle
\end{equation}
which contradicts (\ref{pairprop}).
It can be easily seen that the representation of $S_2$ compatible with the new structure introduced by the twist is now given by $\pi(s_1)\equiv\mathcal{F}_{\theta}\sigma\mathcal{F}^{-1}_{\theta}=\sigma_{\theta}$.  In the language of category theory the new ``flip map" turns the category of Hilbert spaces $\mathcal{H}_{\theta}$ into a ``symmetric braided category", the braiding $\sigma_{\theta}$ being ``symmetric" because $\sigma_{\theta}^2=1\otimes1$.\\ 
The connection between a twisting of the algebra of translations in Minkowski space and the Moyal plane was first discussed in \cite{Watts:1999ra, oeckl-twist}.  In \cite{oeckl-twist}  the extension of the twist to the full Poincar\'e algebra was also discussed but it was only recently that the proposal of a deformed $\theta$-Poincar\'e was re-discovered and gained wider popularity \cite{Chaichian:2004za}. The transformations generated by the elements of the $\theta$-Poincar\'e algebra by construction preserve the commutation relations (\ref{moyalcomm}).  For a comprehensive review of the ``state of the art" for quantum fields on the Moyal plane and their $\theta$-statistics we refer the reader to \cite{akofor}.\\


\section{$\k$-Poincar\'e algebra and deformed Fock space constructions}

As we discussed above in the ``twisted" framework of the Moyal plane only the quasi-triangular structure of the Hopf algebra of the translation generators is changed after deformation.  A natural generalization would be a deformation endowing $U(T_4)$ with a non-trivial {\it coalgebra} structure namely such that the new coproduct $\Delta$ becomes {\it non-cocommutative}, i.e. $\sigma\circ\Delta\neq\Delta$.  Examples of such deformations can be encountered in the translation sector of certain $\kappa$-deformations of the Poincar\'e algebra which we discuss below. 

\subsection{Massless scalar field}
Before embarking in the discussion of the non-trivial coalgebra structure of $\kappa$-deformed tranlsation generators a brief description of the (trivial) coalgebra structure of $U(T_4)$ is in order.  Besides the coproduct for the $P_{\mu}$'s which we introduced in section II
\begin{equation}
\Delta P_{\mu}=P_{\mu}\otimes 1+ 1\otimes P_{\mu}
\end{equation}
the Hopf algebra $U(T_4)$ is equipped with two additional maps the {\it co-unit} $\epsilon:U(T_4)\rightarrow\mathbb{C}$ and the {\it antipode} $S:U(T_4)\rightarrow U(T_4)$
\begin{equation}
\epsilon(P_{\mu})=0\,,\,\,\,\,S(P_{\mu})=-P_{\mu}\, .
\end{equation}
The translation sector of the $\kappa$-Poincar\'e algebra \cite{luk-94}  in the ``bicrossproduct" basis \cite{Majid-Ruegg}, which we denote from now on with $U_{\kappa}(T_4)$, provides a prototypical example of deformation of the trivial coalgebra structure above.  The coalgebra sector of $U_{\kappa}(T_4)$ in fact is given by
\begin{eqnarray}
\Delta(P_0)&=&P_0\otimes 1+1\otimes P_0\,\,\,\,\,\,\Delta(P_j)=P_j\otimes 1+e^{-P_0/\kappa}\otimes P_j\label{coprod}\\ 
S(P_0)&=&-P_0 \\
S(P_l)&=&-e^{\frac{P_0}{\kappa}}P_l \\
\epsilon(P_{\mu})&=&0\, .
\end{eqnarray}
The full $\kappa$-Poincar\'e algebra $\mathcal{P}_{\kappa}$ in the bicrossproduct basis has undeformed Lorentz sector (seen as a projected subalgebra) and and a deformed co-product for boost generators
\begin{equation}
\Delta(N_j)=N_j\otimes 1+e^{-P_0/\kappa}\otimes N_j+\frac{\epsilon_{jkl}}{\kappa}P_k\otimes M_l\, ,
\end{equation}
which also reflects in the non-trivial adjoint action of the boost generators on $U_{\kappa}(T_4)$
\begin{equation}
ad_{N_j} P_l=i\delta_{lj}\Big( \frac{\kappa}{2}  \left(1-e^{-\frac{2 P_0}{\kappa}} \right) +\frac{1}{2 \kappa} \vec{P}^2 \Big)-  \frac{i}{\kappa}P_l P_j\ .
\end{equation}
Using a dual space construction similar to the one we sketched in the previous section for $\mathcal{H}_{\theta}$ and$U_{\theta}(T_4)$ it was shown \cite{Majid-Ruegg} that $U_{\kappa}(T_4)$ is paired with the non-commutative algebra of functions on $\kappa$-Minkowski non-commutative space-time
\begin{equation}
\label{kappaM}
[x_i,t]=\frac{i}{\kappa}x_i\,\,\,\,\,\,\,\,\,\,[x_i,x_j]=0\,\,.
\end{equation}\\
The (deformed) mass Casimir $C_{\kappa}$ of $\mathcal{P}_{\kappa}$
\begin{equation}
\label{casimir}
C_{\kappa}=\left( 2\kappa \sinh \left( \frac{P_0}{2 \kappa}\right)  \right)^2-\vec{P}^2e^{\frac{P_0}{\kappa}}.  
\end{equation}
can be used to construct deformed relativistic field equations \cite{Ruegg:1994bk}. A one-particle Hilbert space carrying an irreducible representation of $\mathcal{P}_{\kappa}$ was first constructed in \cite{AM-fock} for a massless scalar field (massless meaning obviously $C_\k=0$) using the induced symplectic structure form the space of solutions of the undeformed Klein-Gordon equation (see also \cite{Arzano:2007gr} for more details on such construction).  It is clear from the beginning that, due to the asymmetric structure of the deformed coproduct (\ref{coprod}), the flip map is no longer an intertwiner for representations of  $U_{\kappa}(T_4)$ and therefore in the construction of the Fock space a different kind of ``symmetrization", compatible with (\ref{coprod}) must be introduced.  This new ``$\kappa$-symmetrization" is easily illustrated for the simplest case of two-particle states.  Consider for example the
states $|\vec{p}_1\rangle\otimes|\vec{p}_2\rangle$ and $|\vec{p}_2\rangle\otimes|\vec{p}_1\rangle$. Unlike the undeformed case they now have two {\it different} eigenvalues of the linear momentum, respectively $\vec p_1+ e^{-\om(\vec p_1)/\kappa} \vec p_2$ and $\vec p_2+ e^{-\om(\vec p_2)/\kappa} \vec p_1$. Clearly the usual ``symmetrized" two-particle state
\begin{equation}
1/\sqrt{2}(|\vec{p}_1\rangle\otimes
|\vec{p}_2\rangle+|\vec{p}_2\rangle\otimes |\vec{p}_1\rangle)
\end{equation}
is no longer an eigenstate of the momentum operator.  In our context given the two modes $\vec{p}_1$ and $\vec{p}_2$, we have two different $\kappa$-symmetrized two-particle states 
\begin{eqnarray}
|p_{1}p_{2}\rangle _{\kappa } &=&\frac{1}{\sqrt{2}}\left[ |\vec{p}%
_{1}\rangle \otimes |\vec{p}_{2}\rangle +|(1-\epsilon _{1})\vec{p}%
_{2}\rangle 
\otimes  \,|(1-\epsilon _{1}(1-\epsilon _{2}))^{-1}\vec{p}%
_{1}\rangle \right]   \nonumber
\end{eqnarray}%
\begin{eqnarray}
|p_{2}p_{1}\rangle _{\kappa } &=&\frac{1}{\sqrt{2}}\left[ |\vec{p}%
_{2}\rangle \otimes |\vec{p}_{1}\rangle +|(1-\epsilon _{2})\vec{p}%
_{1}\rangle \otimes  \,|(1-\epsilon _{2}(1-\epsilon _{1}))^{-1}\vec{p}%
_{2}\rangle \right]   \nonumber ,
\end{eqnarray}
where $1-\epsilon_i=1-\frac{|\vec{p}_i|}{\kappa}=e^{-\om(\vec p_i)/\k}$.\\
Such symmetrization can be generalized to $n$-particle states in terms of $n$ generators $\pi(s_i)$ belonging to a non-standard representation of the symmetric group $S_n$.  A symmetrized $n$-particle state is obtained from an unsymmetrized one upon action of the operator (\ref{symm}) once the operators $\pi_0(s_i)$ are replaced by operators $\pi(s_i)$ defined by their action on $n$-fold tensor products of single particle states in the following way 
\be \label{kperm} \pi(s_i)|\vec p_1\ra \otimes ... |\vec p_i\ra\otimes |\vec p_{i+1}\ra
\otimes ...|\vec p_n\ra= |\vec p_1\ra \otimes ... |(1-\epsilon _{i})\vec p_{i+1}\ra\otimes |(1-\epsilon _{i+1}(1-\epsilon _{i}))^{-1}\vec p_i\ra \otimes
 ...|\vec p_n\ra\ .
\ee 
It is a bit lengthy but straightforward calculation to show that the relations (\ref{symgroup1}) to (\ref{symgroup3}) hold.  It should be noted that the one-particle states appearing in the tensor products above are on-shell according to the deformed Casimir (\ref{casimir}).  As we will discuss below it seems that this feature must be given up when trying to extend the Fock space construction to massive fields or other bases of the $\kappa$-Poincar\'e algebra.  A salient feature of the deformed Fock space construction is that the two different states $|p_1p_2\rangle_{\kappa}$ and $|p_2p_1\rangle_{\kappa}$ are actually orthogonal as it is easily verified taking the inner product
\begin{equation}
\langle p_1p_2\,|\,p_2p_1\rangle_{\kappa}\simeq\frac{1}{2}
\delta^{(3)}\left(\epsilon_2\vec{p}_1\right)
\delta^{(3)}\left((\epsilon_1(1-\epsilon_2)^{-1}-1)^{-1}\vec{p}_2\right)+
1\leftrightarrow 2\ .
\end{equation}
In general starting from $n$-different modes of the field one will obtain $n!$ orthogonal symmetrized states, one for each eigenvalue of the total momentum\footnote{Note how this differs radically from the $\theta$-deformed case in which the $\theta$-symmetrized states are parallel (see e.g. \cite{akofor})}. This ``fine" structure and the related entanglement between the modes of the field can give rise to interesting phenomena and in \cite{Arzano:2008yc} have been proposed as a possible tool for ``disentangling" the quantum state of the radiation emitted from a black hole 
suggesting a way out of the information paradox.\\
The symmetrized states above were obtained in \cite{AM-fock} looking for tensor product states with the same  eigenvalue for spatial translation generator.  Looking back to the case of quantum fields on the Moyal plane it is natural to ask if there exist some deeper mathematical structure behind the construction, for example, whether or not we can associate a quasitriangular structure to the $\kappa$-deformation of $U(T_4)$ described above.  This a rather non-trivial question given the fact that $U_{\kappa}(T_4)$ as a sub-Hopf algebra of $\mathcal{P}_{\kappa}$ unlike $U_{\theta}(T_4)$ has not been constructed via twisting.  In the next section we address this question in detail and see how it is related to the generalization of the deformed Fock space construction to massive fields and other bases of the $\kappa$-Poincar\'e algebra.

\subsection{A deeper look at $\kappa$-symmetrization}
Attempts to generalize the above construction to a massive scalar field have been unsuccessful so far.  The main reason for this is that the requirement for the vectors in the $\kappa$-symmetrized tensor product (\ref{kperm}) to be on-shell seems to be too restrictive.  If one looks at the multiparticle states of the deformed theory just as generic elements of a Hilbert space, not necessarily decomposable as ordinary symmetrized tensor products of one-particle states then such requirement is not justified as the energy-momentum spectrum of the theory must be on-shell just for the one particle sector of the theory, even in the undeformed case.  A non-standard ``bosonization" procedure can be interpreted, as we will discuss below, as the introduction of a deformed symmetric tensor product $\otimes^s_{\kappa}\equiv\otimes^s\mathcal{R}$ which carries all the structure of the deformed Fock space\footnote{Derivations of $\kappa$-deformed symmetrizations using other strategies have recently appeared in the literature on $\kappa$-field theories see e.g. \cite{Bu-twist, dask, indocroati} }.\\
To be more specific, it can be easily shown that a modification of (\ref{kperm}) leads to a representation of the $S_n$ generators 
\be \label{kperm-ofs} \pi(s_i)|\vec p_1\ra \otimes ... |\vec
p_i\ra\otimes |\vec p_{i+1}\ra \otimes ...|\vec p_n\ra= |\vec p_1\ra \otimes ... |(1-\epsilon_i)\,\vec
p_{i+1}\ra\otimes |(1-\epsilon_{i+1})^{-1}\vec p_i\ra \otimes
 ...|\vec p_n\ra\ ,
\ee 
which is an intertwiner of $U_{\kappa}(T_4)$.  The action of $\pi(s_i)$ in (\ref{kperm-ofs}) above gives again a trivial representation of $S_n$ in which however the field modes are intertwined in a specific way as in the symmetrizazion described in the previous section.  In the present case however the energies of the elements of the tensor product above are unaffected (apart from being permuted along with the state vector of course) and hence
the resulting states are off-shell (for example the state with vector label $(1-\epsilon_i)\vec p_{i+1}$ has
an energy label $\om(\vec p_{i+1})$ rather than $\om((1-\epsilon_i)\vec p_{i+1})$ as it should be if it were
on-shell). It is now very easy to check that the relations (\ref{symgroup1}) to (\ref{symgroup3}) hold, and that after rewriting $1-\eps=e^{-\om(\vec p)/\k}$
they do so independently of the functional form of $\om(\vec p)$, in particular independently of the mass of the
particles.  It is also easy to check that the action of $\pi(s_i)$ commutes with the action of $\D(P_i)$ independently of
the functional form of $\om(\vec p)$. We can therefore use the definition (\ref{kperm-ofs}) in (\ref{symm}) and
proceed to construct the Fock space for the general case on the lines of the massless case recalled in the
previous section.\\ For different bases of the $\kappa$-Poincar\'e algebra, and thus different forms of the coproduct,  one must have different definitions of the representation (\ref{kperm-ofs}).  Below we will show how a general form of such a representation can be found within a more general mathematical framework inspired by the $\th$-Poincar\'e case.
The symmetrization given in (\ref{kperm-ofs}) can indeed be written in a suggestive form if one introduces the operator 
\be \label{twist-p}
 \FF=e^{\frac{1}{\k}P_0\otimes P_j \frac{\p}{\p P_j}}\ . 
\ee 
The action of this operator on a two-particle state is
\be \label{twist-state}
 \FF |\om(\vec p),\vec p\ra\otimes |\om(\vec q),\vec q\ra=|\om(\vec p),\vec p\ra\otimes |\om(\vec q),e^{\om(\vec p)/\k}\vec q\ra\ ,
\ee
as can be verified by acting on the {\it lhs} with the operators $P_0\otimes 1$, $1\otimes P_0$, $P_j\otimes 1$ and $1\otimes P_j$, and noting
that while the first three commute with $\FF$ for the last one we have
\be
[\FF,1\otimes P_j]=e^{P_0/\k}\otimes P_j\ ,
\ee
so that the state $\FF |\om(\vec p),\vec p\ra\otimes |\om(\vec q),\vec q\ra$ is still an eigenstate for such operators,
with the corresponding eigenvalues defining the labels of the state as on the {\it rhs} of (\ref{twist-state}).
We can generalize the operator $\FF$ for an $n$-particle state as
\be \label{twist-pij} 
\FF_i=\underbrace{1\otimes ... 1}_{i-1} \otimes
\FF \otimes \underbrace{1\otimes ... 1}_{n-i-1}\ ,
\ee 
and we can use it to re-express the operators in (\ref{kperm-ofs}) as 
\be \label{twisted-s}
\pi(s_i)=\FF_i \pi_0(s_i)
\FF_i^{-1}\ ,
\ee 
that is, as a twisted representation of the ordinary permutation.
It is now straightforward to show that such a twisted representation of the $S_n$ generators commutes with the
coproduct of momenta in $\k$-Poincar\'e if we recognize that the latter can also be written as a twisted
coproduct. Indeed we have 
\be
 \D(P_\m)=\FF \D_0(P_\m) \FF^{-1}\ ,
\ee 
and  for $n>2$ 
\be 
\D^{n-1}(P_\m)\equiv
(\D\otimes \underbrace{id\otimes ...id}_{n-2})\D^{n-2}(P_\m)=\FF_{12...n} \D_0^{n-1}(P_\m) \FF^{-1}_{12...n}
\ee
where
\be
\begin{split}
& \FF_{12}=[\FF\otimes 1] [(\D_0\otimes id)\FF] \\
& \FF_{123}=[\FF\otimes 1\otimes 1]  [(\D_0\otimes id\otimes id)\FF_{12}] \\ 
& ...
\end{split}
\ee
While in the two-particle case it is trivial to check the commutation $[\D(P_\m),\pi(s_1)]=0$, for $n>2$ one has to notice that (\ref{twisted-s}) can be rewritten as $\pi(s_i)=\FF_{12...n} \pi_0(s_i) \FF^{-1}_{12...n}$ for any $i=1,...,n$.\\ It must be pointed out that the operator $\FF$ does not belong to  $U(T_4)\otimes U(T_4)$ and so it should be regarded as a map on the Hilbert space on which  $U(T_4)$ is acting rather than as a standard twist, despite the similarity of the formulas\footnote{We might elevate $\FF$ to a standard twist by extending the algebra  $U(T_4)$, for example introducing the conjugate operators $X_\m=i\frac{\p}{\p P_{\m}}$ by the Heisenberg double construction \cite{Heis-double}, or by considering it as a twist on the Hopf algebra of $IGL(4,R)$ as in \cite{Bu-twist}. We will not dwell on this here.}.  Keeping in mind the same caveat we can also introduce a quantum R-matrix type of operator, that is, we define the operator
\be \label{R-matrix}
\RR=\FF_{21}\FF^{-1}=e^{-\frac{1}{\k}P_0\wedge P_j \frac{\p}{\p P_j}}\ ,
\ee
where as usual  $a\wedge b\equiv a\otimes b-b\otimes a$, which is such that relation (\ref{r-struct1}) holds. To implement relation (\ref{r-struct2}) we would have to know
what the coproduct of $P_j\frac{\p}{\p P_j}$ is, for which we would have to extend our Hopf algebra. For our purposes it is enough to notice that (\ref{r-struct2}) is a sufficient but
not necessary condition for such operator to satisfy the quantum Yang-Baxter equation $\RR_{12}\RR_{13}\RR_{23}=\RR_{23}\RR_{13}\RR_{12}$,
which indeed is trivially satisfied by (\ref{R-matrix}) because $[P_0,P_j\frac{\p}{\p P_j}]=0$.
Furthermore we have that $\RR^{-1}=\RR_{21}$, which together with the quantum Yang-Baxter equation gives us another demonstration of the fact that our twisted permutations
(\ref{twisted-s}) are a realizations of $S_n$, once we rewrite them as
\be \label{R-nparticles}
\pi(s_i)=\pi_0(s_i)\RR_{i i+1}\ .
\ee
Another advantage of this ``off-shell" construction is that we can relatively easily write a consistent algebra of creation and annihilation operators, $a^\dag_{\vec p}$ and $a_{\vec p}$ respectively.
As usual we define a vacuum state $|0\ra$ such that  $a^\dag_{\vec p}|0\ra = |\vec p\ra$ and $a_{\vec p} |0\ra=0$.
Next we have by construction that
\begin{equation}
a^\dag_{\vec p}\, a^\dag_{\vec q}|0\ra \equiv 1/\sqrt{2} (1+\FF_1\pi_0(s_1)\FF_1^{-1}) |\vec p\ra\otimes|\vec q\ra \equiv |\vec p\ra\otimes^s_{\kappa}|\vec q\ra\, ,
\end{equation}
which can be iterated to the creation of arbitrary multiparticle states.  Note how using such construction one can avoid any reference to ``off-shell" states, {\it the full Fock space is simply constructed using on-shell states and the deformed symmetric tensor product $\otimes^s_{\kappa}$}.  The definition of annihilation operators follows straightforwardly imposing that they annihilate the left-most state in a tensor product (see e.g. \cite{AM-fock}).  We can thus write down  the ``braided" commutators
\begin{eqnarray} 
\label{a+a+}
a^\dag_{\vec p}\ a^\dag_{\vec q}-\RR^{-1}a^\dag_{\vec q}\ a^\dag_{\vec p}&=&0\\
\label{aa}
a_{\vec p}\ a_{\vec q}-\RR a_{\vec q}\ a_{\vec p}&=&0\\
\label{aa+}
a_{\vec p}\ a^\dag_{\vec q} - \FF_{21}\FF a^\dag_{\vec q}\ a_{\vec p}&=&\d_{pq}\ .
\end{eqnarray}
The relations (\ref{a+a+}-\ref{aa+}) are similar to the relations found in \cite{dask} but we have re-derived them here in a different way which highlights the underlying twist structure of the momentum
sector of $\k$-Poincar\'e, and with the help of the R-matrix (\ref{R-matrix}) we have cast them in a form which resembles the case of q-deformed boson algebras, see for example \cite{jeug}.\\
In analogy with the Moyal plane case we can introduce a $\star$-product via the definition $m_{\k}=m\circ\mathcal{F}^{-1}$.  Such product, after substituting $P_{\m}\to -i\frac{\p}{\p x_{\m}}$ and $i\frac{\p}{\p P_{\m}}\to x_{\m}$, leads to $\k$-Minkowski space.  To see this recall that by definition \cite{Majid-Ruegg} $\k$-Minkowski spacetime is the dual space to the translation sector of $\k$-Poincar\'e.
Because of the non-cocommutativity of the latter $\k$-Minkowski turns out to be a non-commutative spacetime with coordinates $\hat x^{\mu}$ satisfying the relations\footnote{As usual Greek indices run from 0 to 3 while Latin indices run from 1 to 3, and repeated indices are summed over. Note that signs and factors of $i$ are not always consistent in the literature.}
\be
[\hat x^0,\hat x^j]=\frac{i}{\k}\hat x^j\ , \hspace{1cm} [\hat x^i,\hat x^j]=0\ ,
\ee
which can be summarized as
\be \label{kMink}
[\hat x^{\m},\hat x^{\n}]=\frac{i}{\k}(\d^{\m}_0\d^{\n}_j-\d^{\n}_0\d^{\m}_j)\hat x^j\ .
\ee
These relations can be realized via the star product $m_{\k}$ applying it to products of the special functions
$f_{\m}(x)=x_{\m}$.
With the explicit form (\ref{twist-p}) we find
\be 
x^{\m}\star x^{\n}=x^{\m}x^{\n} +\frac{i}{\k}\d^{\m}_0\d^{\n}_j x^j\, , 
\ee
which reproduces the relation (\ref{kMink}) above.  This realization can be generalized in the following way\footnote{See also \cite{indocroati} for a related study on twist and $\star$-product on $\k$-Minkowski.}
: if we split the order $1/\k$ part of $x^{\m}\star x^{\n}$
in symmetric and antisymmetric parts we see that the symmetric part does not contribute to the commutators (\ref{kMink}) and hence
can be freely modified without affecting the commutators of $\k$-Minkowski spacetime. For example we can take the following ansatz
\be  \label{star-prod}
x^{\m}\star x^{\n}=x^{\m}x^{\n} +\frac{i}{2\k}(\d^{\m}_0\d^{\n}_j-\d^{\n}_0\d^{\m}_j)x^j +\b \frac{i}{2\k}(\d^{\m}_0\d^{\n}_j+\d^{\n}_0\d^{\m}_j)x^j\ ,
\ee
with $\b$ a free parameter, and with corresponding twist
\be \label{twist1-x}
\FF^{-1}=\exp\left\{ \frac{i}{2\k}\left(\frac{\p}{\p x^0}\otimes x^j\frac{\p}{\p x^j} -
 x^j\frac{\p}{\p x^j}\otimes\frac{\p}{\p x^0}\right) +
 \b \frac{i}{2\k}\left(\frac{\p}{\p x^0}\otimes x^j\frac{\p}{\p x^j} +
 x^j\frac{\p}{\p x^j}\otimes\frac{\p}{\p x^0}\right) \right\}\ .
\ee
It is instructive to look at how the map $\FF^{-1}$ acts on the Fourier basis of plane waves, for which we find that
\be \label{waves}
m(\FF^{-1}\rhd (e^{ip_{\m}x^{\m}}\otimes e^{iq_{\m}x^{\m}}))=\exp\left\{
-i(p^0+q^0)x^0+i(p^j\ e^{\frac{1-\b}{2\k}q^0}+q^j\ e^{-\frac{1+\b}{2\k}p^0})x^j\right\}\ ,
\ee
where we recognize the new addition law for momenta
\be \label{dotplus}
p^\m\dot + q^\m = \begin{cases}  p^0+ q^0 & \hspace{.4cm} \text{if $\m=0$,}\\ p^j\ e^{\frac{1-\b}{2\k}q^0}+q^j\ e^{-\frac{1+\b}{2\k}p^0} & \hspace{.4cm}\text{if $\m=j$.}  \end{cases}
\ee 
Such family of deformed addition laws can be traced back to an ordering ambiguity for functions on $\k$-Minkowski space generalizing what was noted in \cite{agostini}
for the specific cases $\b=1$ and $\b=0$.
We can indeed choose many different orderings for plane waves, in particular the one-parameter family
\be \label{addition}
\vdots e^{ip_{\m}\hat x^{\m}}\vdots_{\b}\equiv
e^{-i\frac{1-\b}{2}p^0\hat x^0}e^{ip^j\hat x^j}e^{-i\frac{1+\b}{2}p^0\hat x^0}\ .
\ee
Products of plane waves ordered in this manner can be recast in the form of a plane wave with the same ordering
and momentum given by the deformed addition law in (\ref{dotplus}).\\
We recognize in (\ref{dotplus}) the non-commutative addition of momenta of $\k$-Poincar\'e in a one-parameter
family of different bases, $i.e.$
\be
\D(P_0)=P_0\otimes 1+1\otimes P_0
\ee
\be
\D(P_j)=P_j\otimes e^{\frac{1-\b}{2\k}P_0}+e^{-\frac{1+\b}{2\k}P_0}\otimes P_j\ ,
\ee
which can be obtained using (\ref{twist1-x}) as a twist, illustrating the link between $\star$-product and deformed coproduct in full analogy with the Moyal plane case
We can now construct a symmetrization for the Fock space following the same steps as in the special case $\b=1$ above just by replacing everywhere the twist operator with the more general one (\ref{twist1-x}),
accomplishing this way the second part of our task, that of generalizing the symmetrization to a whole family of different basis.  Moreover we note that a nice consequence of the definition (\ref{twisted-s}) is that {\it the symmetrization is independent of $\b$}.  Indeed if one rewrites (\ref{twisted-s}) as $\pi(s_i)=\pi_0(s_i)(\FF_i)_{21}\FF_i^{-1}$ it is easy to check that the symmetric part of the twist operator cancel out.


\subsection{$\k$-covariant statistics and the role of the classical $r$-matrix}
Having so far restricted our attention to the translation sector of the $\k$-Poincar\'e algebra the natural question which arises is how the deformed symmetrization used to construct the bosonic Fock space will behave under the action of the full algebra of generators.  Since the deformation that leads to $\k$-Poincar\'e is not a simple Drinfeld twist, as in the Moyal plane case, the compatibility of the ``twist" operator introduced in the previous section with the full deformed algebra must be checked separately.  In fact one can observe that the realization (\ref{twisted-s}) of $S_n$ does not commute with the coproduct of the boost generators, $i.e.$ is not an intertwiner for the whole coalgebra of $\k$-Poincar\'e. This is evident from the expression (\ref{R-nparticles}) since the  operator (\ref{R-matrix}) does not satisfy (\ref{r-struct1}) for the boost sector.  One could accept this as a particular feature of Fock space constructions based on the representations of the $\k$-Poincar\'e algebra, ultimately due to the non-standard deformation used to define the latter, or look for a different construction in which the covariance of the symmetrization under the full $\k$-Poincar\'e algebra is taken as a starting point to build a deformed realization of $S_n$.\\
The first step in the attempt of building a covariant $\kappa$-symmetrization would be to make use of the classical $r$-matrix of $\k$-Poincar\'e (see \cite{r-matrix})
\be
r\equiv i(N_j\otimes P_j - P_j\otimes N_j)\ ,
\ee
where
\be
N_j=i \left( x^0\frac{\p}{\p x^j} + x^j\frac{\p}{\p x^0}\right)
\ee
\be
P_j=-i\frac{\p}{\p x^j}\ ,
\ee
$i.e.$ the usual boost and translation generators, in order to define a new twist map
\be \label{twist2-x}
\FF'=1\otimes 1 + \frac{1}{\k}\left\{ \frac{i}{2}\left(N_j\otimes P_j -
 P_j\otimes N_j\right) +
 \g \frac{i}{2}\left(N_j\otimes P_j +
 P_j\otimes N_j\right) \right\} + O(\frac{1}{\k^2})\ .
\ee
In terms of $\star$-products the map above will correspond to a different choice of the symmetric part in (\ref{star-prod}), $i.e.$
\be \label{t-dependent}
x^{\m}\star x^{\n}=x^{\m}x^{\n} +\frac{i}{2\k}(\d^{\m}_0\d^{\n}_j-\d^{\n}_0\d^{\m}_j)x^j +\b \frac{i}{2\k}(\d^{\m}_0\d^{\n}_j+\d^{\n}_0\d^{\m}_j)x^j +\g \frac{i}{\k}\d^{\m}_j\d^{\n}_j x^0\ ,
\ee
for the special case $\b=\g$.
The use of such twist function has the main advantage of being automatically covariant under action of the full $\k$-Poincar\'e algebra, as the coproduct of the boost generators
is also modified with the same twist (to order $1/\k$ at least).  Note how, as before, the choice of the symmetric part in (\ref{t-dependent}) dictates only linear order terms in $1/\kappa$ in the definition of the twist map and this leaves us with a freedom in choosing $\FF'$. Unlike the case of (\ref{twist1-x}), where the expression of the twist to all orders was inferred from the coproduct structure of the momentum sector, now the difficulty lies in reconstructing the higher order terms and thus the corresponding coproduct \footnote{Using (\ref{twist2-x})  for $\g=0$ as a twist we obtain to order $1/\k$ the coproduct of $\k$-Poincar\'e in the symmetric basis ($\b=0$), this is not true at higher orders or for other values of $\g$.}.\\
A naive extension of this construction to all orders in $1/\k$ through exponentiation of the linear order term in (\ref{twist2-x}) does not generate the $\k$-Poincar\'e algebra.  In fact, as noted recently in \cite{young3}, such construction does not exist.
This would not be a problem in general because for a non-cocommutative Hopf algebra being a twist of a cocommutative one is only a sufficient condition for the existence of an intertwiner, not a necessary one.
A more general sufficient condition is that the Hopf algebra is a quasi-triangular one, in which case one can define the intertwiner directly from the quantum R-matrix as in (\ref{R-nparticles}),
but it is not known if $\k$-Poincar\'e has a quasi-triangular structure, and in particular the R-matrix of $U_q(so(3,2))$ diverges in the contraction limit that leads to $\k$-Poincar\'e \cite{luk-94}.\\
The idea of exploiting the R-matrix underlies the attempt of  Young and Zegers \cite{young} in defining a $\k$-Poincar\'e covariant statistics. Indeed
by explicit computation one finds (as noted in \cite{young3} where the computation of the R-matrix is done to order $1/\k^3$) that at second order in $1/\k$ the symmetrization in \cite{young} is given by
\be \label{R-2particles}
\pi'(s_1)=\pi_0(s_1)\RR'
\ee
where
\be
\RR'=e^{ \frac{1}{\k}(\bar N_j\otimes\bar P_j -\bar P_j\otimes \bar N_j)}+O(\frac{1}{\k^3})
= e^{ \frac{1}{\k}(( N_j+\frac{\eps_{jlm}}{2\k}M_l P_m)e^{\frac{P_0}{2\k}}\wedge  P_j e^{\frac{P_0}{2\k}})}+O(\frac{1}{\k^3})\ .
\ee
which also satisfies (\ref{r-struct1}) at least to order $1/\k^2$ for all the generators of $\k$-Poincar\'e.
Note that we have introduced here the generators of the original (sometime called ``standard") basis \cite{luk-94} of $\k$-Poincar\'e  $\{\bar P_\m,\bar N_j,\bar M_j\}$
linked to the bicrossproduct basis by the transformations
\be
P_0 =\bar P_0\ , \hspace{.2cm} P_j = \bar P_j e^{-\frac{\bar P_0}{2\k}}\ , \hspace{.2cm} 
N_j= \bar N_j e^{-\frac{\bar P_0}{2\k}}-\frac{\eps_{jlm}}{2\k}\bar M_l \bar P_m e^{-\frac{\bar P_0}{2\k}} \ , \hspace{.2cm} M_j=\bar M_j\ .
\ee
If we naively extend (\ref{R-2particles}) to $n>2$ particles as in (\ref{R-nparticles}) we would run into troubles because, as has been known for a long time \cite{r-matrix}, $r$ does not satisfy the classical Yang-Baxter equation
and hence $\RR'$ (at order $1/\k^2$) does not satisfy its quantum counterpart either, implying that also $\pi'(s_i)=\pi_0(s_i)\RR_{ii+1}$ do not satisfy the classical Yang-Baxter equation
(\ref{symgroup2}) and hence $\pi'(s_i)$ do not form a realization of $S_n$. 
A solution to this problem has been proposed in \cite{young3} with the introduction of a coassociator and a quasibialgebra structure, but it is unclear at this stage whether the ambiguity in the symmetrization introduced by such procedure would lead to physically distinguishable states or not.


\section{Conclusions}
We provided an overview of the various types of ``momentum-dependent" statistics encountered in non-commutative quantum field theories based on quantum deformations of the Poincar\'e algebra.  In particular we focused on $\kappa$-deformations of the latter and reviewed different types of $\kappa$-symmetrizations appearing in the literature.  With the use of a ``twist-like" formalism we provided a general framework for extending the deformed symmetrization procedure of \cite{AM-fock} to {\it all} bases of the $\kappa$ Poincar\'e algebra and to massive fields.  The twist-like and the associated R-matrix-type operators introduced allowed us to write down the deformed algebra of creation and annihilation operators and provided a clearer picture of the interplay between different choices of bases in $\kappa$ Poincar\'e and the degeneracy in defining a $\star$-product for non-commutative functions on $\kappa$-Minkowski space.\\ The current open issue in such constructions is the understanding of the covariance properties of the intertwiners for the $\kappa$-deformed momentum sector under the action of the full set of deformed momentum generators.  As it stands the symmetrization given in section turns out to be non-covariant under the action of deformed boosts.  This, on one side, suggests a different approach to the problem namely to try to construct a symmetrization imposing the requirement of covariance from the beginning, as in \cite{young,young3}, but, as we mentioned above, to date results in this direction are only perturbative and seem to be plagued by ambiguities in the multi-particle sector.
On the other hand it could simply indicate that the very notion of covariance for $\kappa$-deformed multi-particle states breaks down if one chooses to describe the $\kappa$-Fock space in terms of plane wave basis vectors in a way reminiscent of the description of the Fock space of local quantum field theory in terms of the non-covariant Newton-Wigner localized states.  Clarifying these issues remains a priority for future work on models of quantum fields with $\kappa$-symmetries.

\begin{acknowledgments}
We would like to thank  Laurent Freidel, Jurek Kowalski-Glikman, Antonino Marcian\'o, Robert Oeckl, Federico Piazza, Simone Speziale and Charles Young for
discussions.   
Research at Perimeter Institute for Theoretical Physics is supported in part by the Government of Canada through NSERC and by the Province of Ontario through MRI.
\end{acknowledgments}

\bibliographystyle{unsrt}

\end{document}